# MULTIFREQUENCY OBSERVATIONS OF GAMMA-RAY BURSTS

JOCHEN GREINER

*Max-Planck-Institute for Extraterrestrial Physics, 85740 Garching, Germany*

**ABSTRACT.** Neither a flaring nor a quiescent counterpart to a gamma-ray burst has yet been convincingly identified at any wavelength region. The present status of the search for counterparts of classical gamma-ray bursts is given. Particular emphasis is put on the search for flaring counterparts, i.e. emission during or shortly after the gamma-ray emission.

## 1. Introduction

There are increasing doubts that the puzzle of $\gamma$-ray bursts (GRBs) will be solved by simply accumulating more events with the gamma-ray detectors currently in operation. While the solution might come with an unpredictable circumstance, the identification of a GRB counterpart at any other wavelength is generally believed to be one of the most obvious possible ways to unveil the enigma. This belief manifests itself in the sudden increase in activity over the last 1–2 years in all major wavelength bands.

Counterpart searches are hampered by many unknowns. It is not known whether a GRB will cause a permanent change in the emission in one or several wavelength bands, or will yield a multiwavelength transient event. If there is a transient event at other energies correlated with a GRB, it is not known whether it occurs before, during or after the $\gamma$-ray event and how long it might last.

The counterpart search is most promising in a few specific energy bands for very different reasons: First of all, there are the adjacent energy ranges at GeV and soft X-rays where some GRBs have been seen to emit part of their power. But in addition, radio observations seem to combine two other advantages: (1) Basically all known $\gamma$-ray sources are also radio emitters, in most cases even strongly variable and thus easy to discover. (2) If GRBs are indeed at cosmological distances as suggested by their isotropy and non-homogeneity, then the dispersion will delay the signal up to an hour at 25 MHz (Palmer, 1993) and thus help in succeeding in "simultaneous", rapid follow-up observations.

Here, I review the search for GRB counterparts in all wavelength bands. Previous reviews can be found in Schaefer (1994a), Hartmann (1995) and Greiner (1995).

## 2. Search Strategies

The search for simultaneous or near-simultaneous observations of GRBs at other wavelengths can be attacked with two different strategies: (1) either correlating independently conducted monitoring observations which happen to overlap in time or (2) aiming in quick follow-up observations after the onset of a GRB. Both of these approaches have become reasonable and practical only with the launch of the *Compton* Gamma-Ray Observatory (GRO) and the BATSE (Burst and Transient Source Experiment) measurements: The correlation research due to the detection and localization of about one burst per day, and the rapid follow-up observations due to the failure of the GRO tape recorders and the advent of fast, on-line data transmission and reduction.

While the rapid follow-up observations have now set the most stringent limits for emission from GRBs at times greater than 1 day after the burst event, most of the simultaneous limits so far arose from the correlation of BATSE burst locations/times with monitoring observations in various other bands. These include the correlation with

- COBE scanning observations (Bontekoe *et al.*, 1995)
- regularly exposed wide-field photographs of most major observatories (Greiner *et al.*, 1994),
- CCD-based multi-camera sky monitoring of 0.75 ster (Vanderspek *et al.*, 1995),
- atmospheric airshower events due to TeV emission in arrays such as CYGNUS (Alexandreas *et al.*, 1994), HEGRA (Matheis *et al.*, 1995), and Gran Sasso (Aglietta *et al.*, 1995a),
- neutrino or muon detection rates with IMB (LoSecco 1994, Becker-Szendy *et al.*, 1995), Soudan 2 (DeMuth *et al.*, 1994) and Gran Sasso (Aglietta *et al.*, 1995b).

## 3. Search Results

### 3.1. Radio emission

A simultaneous observation of GB 911226 with the COBE satellite was identified by correlating the 207 GRBs which occurred during the 8 month period of overlap with the COBE database (Bontekoe *et al.*, 1995). COBE scanned the sky at the three frequencies 31, 53 and 90 GHz. The COBE field of view (FOV) started covering the 10° radius error box of the rather faint burst GB 911226 about 0.1 s after burst onset, reaching 80% coverage after the next few seconds (the burst duration $T_{90}$ is 1.3 sec). No source was detected yielding an $2\sigma$ upper limit of 31000 Jy at 53 GHz (Bontekoe *et al.*1995).

The error box of GB 920711 was serendipitously observed 2 and 4 days after the $\gamma$-ray event with the Cambridge Low Frequency Synthesis Telescope (CLFST) at 151 MHz. No evidence for variability with respect to earlier observations of this field in 1975 and to later follow-up observations in June/July 1993 have been found within the noise limit of about 40 mJy at 1′ resolution (Koranyi *et al.*, 1994).

The bright gamma-ray burst GB 940301 was monitored at 1.4 GHz (2°.6 FOV) and 0.4 GHz (8°.1 FOV) at the Dominion Radio Astrophysical Observatory (DRAO) Synthesis Telescope starting starting 3 days after the burst (Frail *et al.*, 1994). A total of 245 radio sources were identified in the summed image in an intensity range between 0.8–110 mJy, but none was identified as a candidate for a flaring/fading radio counterpart based on variability analyses. The daily upper limits on the nondetection are ≈3.5 mJy at 1.4 GHz and 55 mJy at 0.4 GHz and allow to constrain some fireball parameters of cosmological GRB models (Frail *et al.*, 1994). GB 940301 was also observed at 92 cm at Westerbork. The first observation of the corrected final position took place 32 days after the burst, and an upper limit of $< 40$ mJy has been derived (Hanlon *et al.*, 1995).

Deep VLA observations of three BATSE GRBs have been performed over the last two years (Palmer *et al.*, 1995). GB 920501 was observed after 9 months at 20 cm and 3.6 cm with upper limits of $< 325$ $\mu$Jy/beam and $< 77$ $\mu$Jy/beam, respectively. Similar limits have been derived at 20 cm for GB 930706 ($< 1$ $\mu$Jy after 9 days) and GB 940217 ($< 4$ $\mu$Jy after 23 hours).

### 3.2. Infrared

The only major search for quiescent infrared emission from GRBs has been performed by checking the IRAS database for 23 well localized error boxes (Schaefer *et al.*, 1987). The three sources found inside the error boxes have been shown to not be related to the GRB. In addition, three error boxes have been observed with the 1.3 m Kitt Peak National Observatory (KPNO) in the 2.2 $\mu$m (K) band, and again no counterpart was found down to a limiting magnitude of K = 13.6.

### 3.3. Optical

#### 3.3.1. Archival search

Historically, the search for GRB counterparts began with the check of photographic exposures taken contemporaneously with the burst event (Grindlay, Wright and McCrosky 1974). Searches for archival optical transients in small GRB error boxes using large photographic plate collections were initiated at Harvard Observatory (Schaefer *et al.*, 1984) and then also performed at other observatories (Hudec *et al.*, 1987; Greiner *et al.*, 1987). Though more than 130 thousand plates have been investigated (see Tab. I) and several optical flash objects were found, no convincing relation could be established between archival optical objects and GRBs. The most convincing optical transient event which was imaged on three simultaneously exposed plates turned out to be a large amplitude flare of a dMe star (Greiner and Motch 1995).

TABLE I
Archival Search for GRB optical counterparts

| Group | Observatories | No. of GRB error boxes | No. of plates | monitoring time (years) |
|---|---|---|---|---|
| Schaefer *et al.* | Harvard | 16 | 32000 | 4.25 |
| Hudec *et al.* | Ondřejov | 21 | 30000 | 10 |
| Greiner *et al.* | Sonneberg | 15 | 35000 | 2.6 |
| Moskalenko *et al.* | Odessa | 40 | 40000 | 1.3 |
| Schwartz *et al.* | Santa Barbara | 7 | (photoelectric) | 0.1 |

*3.3.2. Correlating optical sky patrols with GRBs*

Regular, wide-field sky patrols of two kinds are presently correlated with GRBs detected by BATSE: the Explosive Transient Camera (ETC) exposures with a total FOV of 40°×60° (Vanderspek *et al.*, 1995), and the logistic network of photographic patrols performed at a dozen observatories worldwide (Greiner *et al.*, 1994).

The ETC is a wide (0.75 ster) FOV CCD array consisting of 8 camera pairs which operates entirely by computer control on Kitt Peak National Observatory (Vanderspek *et al.*, 1994). During its more than 4 years of operation there were five cases in which a BATSE GRB occurred during an ETC observation and within or near an ETC FOV (Krimm *et al.*, 1994, Vanderspek *et al.*, 1995). No optical transients were detected during these observations, resulting in upper limits for the **fluence** ratio $L_\gamma/L_{opt} \geq$ 2-120. Unfortunately, in all cases of simultaneous exposures only a part (20%–80%) of the rather large BATSE error box was covered. For those GRBs also seen by other satellites, triangulation will reduce the error box size and may increase the actual coverage.

The correlation of BATSE GRBs and photographic wide-field plates of the network of 11 observatories with on-going regular photographic observations has identified simultaneous plates for nearly 60 GRBs, with typically $m_{lim} \approx$ 2–3 mag for a 1 sec duration flash (Greiner *et al.*, 1994). Blink comparison of these plates has not revealed any optical transient (except new variable stars) resulting in limits for the **flux** ratio $F_\gamma/F_{opt} \geq$ 1-20.

*3.3.3. Rapid follow-up observations*

Rapid follow-up observations rely on BACODINE (BAtse COordinate DIstribution NEtwork) which computes and distributes coordinates of strong GRBs within typically 5 sec to interested observers (Barthelmy and Fishman 1995). For bursts occurring in the FOV of the imaging COMPTEL telescope, coordinates can be determined with much better accuracy and are distributed after typically 15–30 minutes via the BATSE/COMPTEL/NMSU network (Kippen *et al.*, 1994).

From the wealth of observational results I can pick out only one example, which should demonstrate the progress possible with the rapid burst notification: Within the observing campaign of GB 940301 by the BATSE/COMPTEL/NMSU rapid response network the COMPTEL error box was observed seven hours after the burst with the 1m Schmidt telescope at Socorro reaching a limiting magnitude of $m_V \approx$ 16 mag (Harrison *et al.*, 1995). Again, no optical transient was found.

3.4. EUV / Soft X-rays

*Ginga* observations have shown that some GRBs are detectable down to 2 keV. Very recently, the BATSE spectroscopy detectors have successfully searched for GRB emission in the 5–10 keV range (Preece *et al.*, 1995), and WATCH has seen several GRBs to be quite strong in the 6–15 keV range. Unfortunately, apart from these observations there are no other simultaneous detections of GRBs at soft X-rays mainly due to the small FOV of the instruments.

The first search for quiescent X-ray sources in 5 GRB error boxes has been conducted with *Einstein* (Pizzichini *et al.*, 1986) and *EXOSAT* (Boër *et al.*, 1988). Using the *ROSAT* all-sky-survey data this search has been extended to more than 30 GRB error boxes (Greiner *et al.*, 1995a) of the $2^{nd}$ (Atteia *et al.*, 1987) and 15 error boxes of the $3^{rd}$ IPN catalog (Hurley *et al.*, 1993). A total of 27 X-ray sources have been discovered, but this number corresponds to the mean source density in the *ROSAT* survey. The optical identification of these sources has not revealed any unusual associations, and thus none of these X-ray sources is considered as a quiescent GRB counterpart. Depending slightly on the spectral model and the absorption used, the conversion from the upper limit countrate to flux limits results in $\approx 10^{-13}$ erg/cm$^2$/s (0.1–2.4 keV).

Though the optical identification is not yet complete, the up to now discovered *ROSAT* X-ray sources in or very near to a GRB error box do not reveal a systematic correlation between GRB error boxes and any known population. This negative result can be used to constrain some of the existing GRB models with neutron stars as ingredients. Assuming slow neutron stars accreting from the interstellar matter, the thermal emission of the neutron star should be detectable up to a certain distance depending on the emitting area of the neutron star and the absorbing column (Pizzichini *et al.*, 1986; Boër *et al.*, 1988). For the most conservative case of small (polar cap) emitting area and maximum absorption the typical minimum distances are $\sim$200 pc (Boër *et al.*, 1994, Greiner *et al.*, 1995a). Similar distance limits can be derived for the model of Epstein (1985) where sudden infall of $\geq 10^9$ g of matter onto a (slowly rotating, weakly magnetic) neutron star explains the energetics of a GRB. In order to prevent frequent repetitions, the mean mass accretion rate between the bursts has to be very slow suggesting that the disk is rather dense and relatively cool, and assuming blackbody radiation one again obtains the temperature (of the inner disk) as a function of the distance. These are additional arguments against a galactic disk neutron star population as GRB sources.

For GRBs of the $3^{rd}$ IPN the *ROSAT* observations took place a few up to 18 months before the burst event. It has been argued (Lasota, 1992) that under the assumption of a slowly accreting neutron star as a $\gamma$-ray burst source, the burst could produce a shock which would prevent accretion onto the neutron star for a time span of several years following the burst. The upper limits on the X-ray flux constrain the pre-burst accretion rate of the neutron star and thus, only for distances larger than a few hundred pc (see Greiner *et al.*, 1995b for one specific example) would the accretion rate be high enough to trigger a hydrogen flash (Hameury *et al.*, 1983).

Attempts for rapid follow-up observations of well localized GRBs with *ROSAT* have been made for GB 920501 (Li *et al.*, 1995) and the suspected "*COMPTEL*" repeater GB 930704/940301 (Greiner *et al.*, 1995c). Both *ROSAT* pointings occurred 4 weeks after the burst event and revealed no fading counterpart. Deep *ROSAT* pointings on a few selected small GRB error boxes have also failed to unveil sources with unusual properties, and the deduced upper limits are $10^{-14...-15}$ erg/cm$^2$/s (Boër *et al.*, 1993).

A deep EUVE pointing on GB 920325 about 17 months after the burst failed to detect a quiescent EUV counterpart. The $3\sigma$ upper limits (Hurley *et al.*, 1995) of $3 \times 10^{-14}$, $1 \times 10^{-12}$, and $5 \times 10^{-13}$ erg/cm$^2$/s in the 40–190, 140–380 and 280–760 Å bands, respectively, impose similar constraints on thermally radiating quiescent counterparts as the above discussed *ROSAT* survey limits.

## 3.5. GeV / TeV Emission

A correlation of the 1991/1992 season 0.4–4 TeV events observed by the 10m Whipple Observatory Reflector with BATSE bursts has resulted in no positive detection which allows to set limits on the density of primordial black holes and the local density of cosmic strings (Connaughton et al., 1994).

The CYGNUS-I air shower array detects ultra-high-energy (UHE) radiation at about 3.5 triggers/sec for E > 50 TeV from zenith and E > 100 TeV for $\theta > 30°$. For 52 out of 260 BATSE GRBs and 6 (out of 18) IPN GRBs having occurred in the CYGNUS FOV ($\theta < 60°$) no evidence for UHE emission was found. The limits for 3 GRBs are inconsistent with an extrapolation of the BATSE spectra, indicating either a softening of the production spectrum at high energies or the presence of UHE $\gamma$-ray absorption. The latter, if true, would imply cosmological GRB distances (Alexandreas et al., 1994).

All GRBs of the BATSE 2B catalog with a zenith angle < 60° and a location error < 15° have been correlated with the HEGRA triggers. There are simultaneous data for 85 bursts but no significant excess could be found around the burst trigger times resulting in an upper limit between $10^{-7}...10^{-11}$ cm$^{-2}$ s$^{-1}$ for energies in the range of 40–500 TeV (Matheis et al., 1995). A HEGRA search in a time window of 12 hours after the bursts also revealed no significant excess rate. This excludes strong extended TeV emission, as one could expect from the observation of the extended GeV emission from GB 940217 (Hurley et al., 1994).

Further correlations with BATSE bursts with the Thebarton air shower array above 100 TeV (Smith et al., 1995) and with the Gran Sasso air shower array (Aglietta et al., 1995a) in the 10–100 GeV range (upper limit of $2\times10^{-4}$–$3\times10^{-3}$ erg/cm$^2$) and above 100 TeV ($3\times10^{-6}$–$1\times10^{-5}$ erg/cm$^2$) also revealed no excess emission.

## 3.6. Neutrinos and muons

Searches for neutrinos, antineutrinos and muons during times of BATSE GRBs have been performed using data of the Irvine-Michigan-Brookhaven (IMB) detector (LoSecco 1994; Becker-Szendy et al., 1995)), the Soudan 2 detector (DeMuth et al., 1994) and the Mont Blanc Neutrino Telescope (Aglietta et al., 1995b). The neutrino and muon samples are believed to contain primarily atmospheric events, i.e. neutrinos produced by the decay of secondary particles from cosmic-ray interactions in the Earth's atmosphere. The IMB correlation with 183 Venera and Ginga GRBs revealed no neutrino coincidences. The resulting flux limit is $5\times10^5$ neutrinos cm$^{-2}$ for neutrinos and antineutrinos in the 60–2500 MeV range (Becker-Szendy et al., 1995). Also, no muon coincidences were found. A flux limit of $5.8\times10^{-7}$ muons cm$^{-2}$ from neutrino interactions during GRBs is derived (LoSecco 1994).

A correlation of 180 BATSE GRBs with the Soudan 2 muon detector events revealed 18 trigger coincidences, near the number expected by chance (DeMuth et al., 1994). A correlation of 553 BATSE GRBs with events of the Mont Blanc telescope revealed no excess interactions. Depending on the neutrino flavour, the upper limits range between $6\times10^{11}$ cm$^{-2}$ (for $\tilde{\nu}_e$ at 9$\leq$E$_\nu$ $\leq$50 MeV), $2.1\times10^{13}$ cm$^{-2}$ (for $\nu_e, \nu_\mu, \nu_\tau$ at 20$\leq$E$_\nu$ $\leq$100 MeV) and $2.2\times10^{13}$ cm$^{-2}$ (for $\tilde{\nu}_e, \tilde{\nu}_\mu, \tilde{\nu}_\tau$ at 20$\leq$E$_\nu$ $\leq$100 MeV) (Aglietta et al., 1995b).

## 4. Summary and Prospects

Despite the increasing efforts in the search for GRB counterparts no convincing candidate has yet been identified. While the limits for simultaneous emission are still not deep enough to make statements on the possible necessity of breaks in the extrapolation of the gamma-ray spectra, the limits for up to a few hours after the bursts are getting constraining for models which predict either (1) strong and/or long-duration afterglows, or (2) simple extrapolations of the measured X-ray to $\gamma$-ray spectra to lower and higher energies (see Fig. 1).

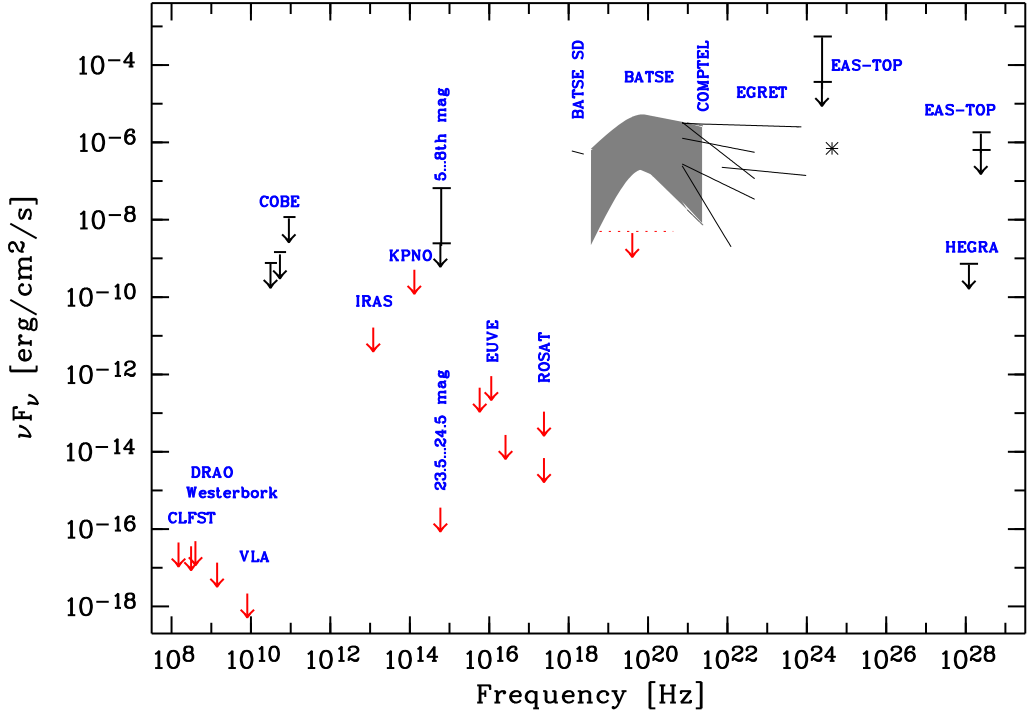

Fig. 1. Upper limits to simultaneous and quiescent emission from various GRBs from radio up to TeV energies. The shaded area marks the range of GRB spectra as seen by BATSE (Band et al., 1993). Straight lines mark the best fit power laws in the 1 MeV to few GeV range of the few bursts with detected emission, and the asterisk marks the 18 GeV photon of GB 940217. Arrows with horizontal bar are simultaneous limits while arrows without bar are from non-simultaneous follow-up observations. Radio: Frail et al.(1994), Koranyi et al.(1994), Hanlon et al.(1995), Palmer et al.(1995). COBE: Bontekoe et al.(1995). IRAS and KPNO: Schaefer et al.(1987). Optical: Motch et al.(1985), Greiner et al.(1994), Vrba et al.(1994), Luginbuhl et al.(1995), Akerlof et al.(1995). ROSAT: Boër et al.(1993), Greiner et al.(1995a). BATSE: Band et al.(1993), Preece et al.(1995), Horack and Emslie (1994). COMPTEL: Hanlon et al.(1994). EGRET: Dingus et al.(1994), Hurley et al.(1994). EAS–TOP: Aglietta et al.(1995a,b). HEGRA: Matheis et al.(1995).

Observational improvements in the very near future can be expected mainly from (1) the upcoming interplanetary network of BATSE, Ulysses, WIND and several other satellites yet to be launched (Hurley 1995), and (2) the launch of HETE (Ricker 1988) which will measure burst positions with better than $10'$ accuracy (see Tab. II) and after subsequent data transmission to ground observers will allow observations with smaller FOV telescopes (and thus fainter limiting magnitudes) than feasible up to now. This for the first time will allow optical observations with a sensitivity which is comparable to flux estimates deduced by extrapolating burst spectra down from the X-ray/$\gamma$-ray range (Band and Ford, 1995).

TABLE II
Parameters for rapid burst position distribution

|  | BACODINE | COMPTEL/BATSE/NMSU | HETE |
|---|---|---|---|
| Time delay | 5 sec | 16 min | 5 sec |
| Location error | $\pm 10°$ | $\pm 1\text{-}2°$ | $\pm 10'$ |
| Rate | 20–50 $yr^{-1}$ | 2–5 $yr^{-1}$ | 5–20 $yr^{-1}$ |

For future developments the following two regions are thought to be most promising: First, monitoring the soft X-ray emission contemporaneous to GRBs would not only allow an improved localization of individual bursts, but would also establish a rough distance scale depending on the variation of absorption in spectra of different GRBs occurring at different galactic latitude (Schaefer 1994b). Second, improving the sensitivity and/or lowering the threshold of air shower experiments in the GeV range would certainly improve our ability of establishing the fraction of bursts with extended high-energy emission.

### Acknowledgements

JG is supported by the Deutsche Agentur für Raumfahrtangelegenheiten (DARA) GmbH under contract FKZ 50 OR 9201.

**Auriemma:** Is there any reason why the optical flash should not come first, perhaps minutes or hours before the GRB?

**Greiner:** As long as we do not know what causes a GRB, and hence do not know the emission process, we cannot rule out the possibility that the optical flash comes earlier than the GRB itself. If this were the case then only wide-field monitoring campaigns can hope to identify this flash. The correlation process of GRBs with wide-field photographic plates has not only focused on simultaneous plates but also identified a number of near-simultaneous plates (Greiner et al., 1994). Due to the relaxed time constraint there are several plates with substantial fainter limiting magnitude (17th = about a factor of 100 deeper than the simultaneous limits) within $\pm 12$ hours, and no flash has been found on these so far.